\documentclass[10pt,twocolumn]{article}

\usepackage[margin=0.75in]{geometry}
\usepackage{amsmath}
\usepackage{amssymb}
\usepackage{graphicx}
\usepackage{booktabs}
\usepackage{xcolor}
\usepackage{hyperref}

\title{\textbf{Hierarchical Preemption: A Novel Information-Theoretic Control Mechanism in Lambda Phage Decision-Making}}

\author{
Eugênio Simão\textsuperscript{1,*}\\
\textsuperscript{1}Department of Computer Science\\
Universidade Federal de Santa Catarina (UFSC)\\
Araranguá, Santa Catarina, 88906-072, Brazil\\
\textsuperscript{*}Corresponding author: eugenio.simao@ufsc.br
}

\date{December 26, 2025}

\begin{document}

\maketitle

\begin{abstract}
Biological systems organize into hierarchies to manage complexity, yet the mechanisms governing hierarchical control remain incompletely understood. Using information theory \cite{shannon1948,cover2006} and the Lambda phage lysis-lysogeny decision as a model system, we discover that hierarchical control operates through \textit{hierarchical preemption}—higher layers collapse decision space rather than blocking lower-layer signals. Through mutual information (MI) analysis of 200 stochastic simulations, we demonstrate that the UV damage sensor (RecA) achieves 2.01× information advantage over environmental signals by preempting bistable outcomes into monostable attractors (98\% lysogenic or 85\% lytic). Conditional MI analysis reveals that the integrator signal (CII) carries lower information when RecA is absent (saturated, 0.06 bits) than when RecA is active (subsaturated, 0.38 bits). This \textit{saturation effect} demonstrates that signals orchestrate compartment behaviors by removing decision space—achieving 85-98\% outcome certainty while preserving 2-15\% escape routes. These findings establish a quantitative framework for hierarchical information processing in cellular decision-making.
\end{abstract}


\section{Introduction}

Biological systems face a fundamental challenge: managing overwhelming complexity while maintaining reliable function. The human cell coordinates ~20,000 genes, ~100,000 proteins, and countless metabolites \cite{alberts2015}—a computational burden that would overwhelm centralized control. Evolution's solution is hierarchical organization: environmental signals compress into integrator signals, which drive binary decisions \cite{alon2007}.

The Lambda bacteriophage lysis-lysogeny decision exemplifies this strategy \cite{ptashne2004}. Upon infecting \textit{E. coli}, Lambda integrates multiple signals—UV damage, nutrient availability, cell cycle state—to choose between immediate lysis (lytic pathway) or dormant integration (lysogenic pathway). Four key regulatory proteins orchestrate this decision: CI and Cro repressors compete for control of the genetic switch through mutual inhibition, where high CI levels establish lysogeny and high Cro levels trigger lysis \cite{ptashne2004,little2014}. CII protein integrates environmental signals to bias transcription toward lysogenic commitment by promoting CI production \cite{kourilsky2016}. RecA protein senses DNA damage (primarily UV-induced lesions) and activates CI degradation, driving the lytic pathway under stress conditions \cite{st-pierre2008,zeng2010}. This binary choice emerges from a multilayer signaling hierarchy involving these four proteins, yet how information flows through this hierarchy remains quantitatively uncharacterized.

\subsection{Classical Hierarchical Control Theory}

Traditional models of hierarchical control assume \textit{signal gating}: higher layers block lower-layer information flow \cite{simon1962,ferrell2012,levchenko2018}. In this framework, a hierarchical override (e.g., UV damage) should suppress subordinate signals (e.g., metabolic state), reducing their predictive power. This predicts:
\begin{equation}
I(\text{CII}; \text{Decision} \mid \text{RecA}_{\text{off}}) \gg I(\text{CII}; \text{Decision} \mid \text{RecA}_{\text{on}})
\end{equation}
where $I(X;Y)$ denotes mutual information (MI) \cite{cover2006}. However, our findings challenge this prediction.

\subsection{Our Discovery: Hierarchical Preemption}

We report the opposite: RecA's hierarchical priority operates through \textit{decision space collapse}, not signal blocking. When RecA is inactive, the integrator CII becomes saturated (all values above threshold), yielding \textit{low} information (0.06 bits) despite mechanistic freedom. When RecA is active, CII becomes subsaturated (spanning threshold), yielding \textit{high} information (0.38 bits) by predicting which cells escape the lytic attractor. This \textit{hierarchical preemption} mechanism reveals that biological hierarchies achieve robustness (85-98\% outcome certainty) without complete signal suppression, preserving flexibility through stochastic escapes (2-15\%).

\section{Model Architecture}

To test these contrasting predictions quantitatively, we built an extended Lambda phage model incorporating hierarchical signal integration with explicit mechanistic detail.

\subsection{Hierarchical Lambda Phage Network}

\begin{figure}[htbp]
\centering
\includegraphics[width=\columnwidth]{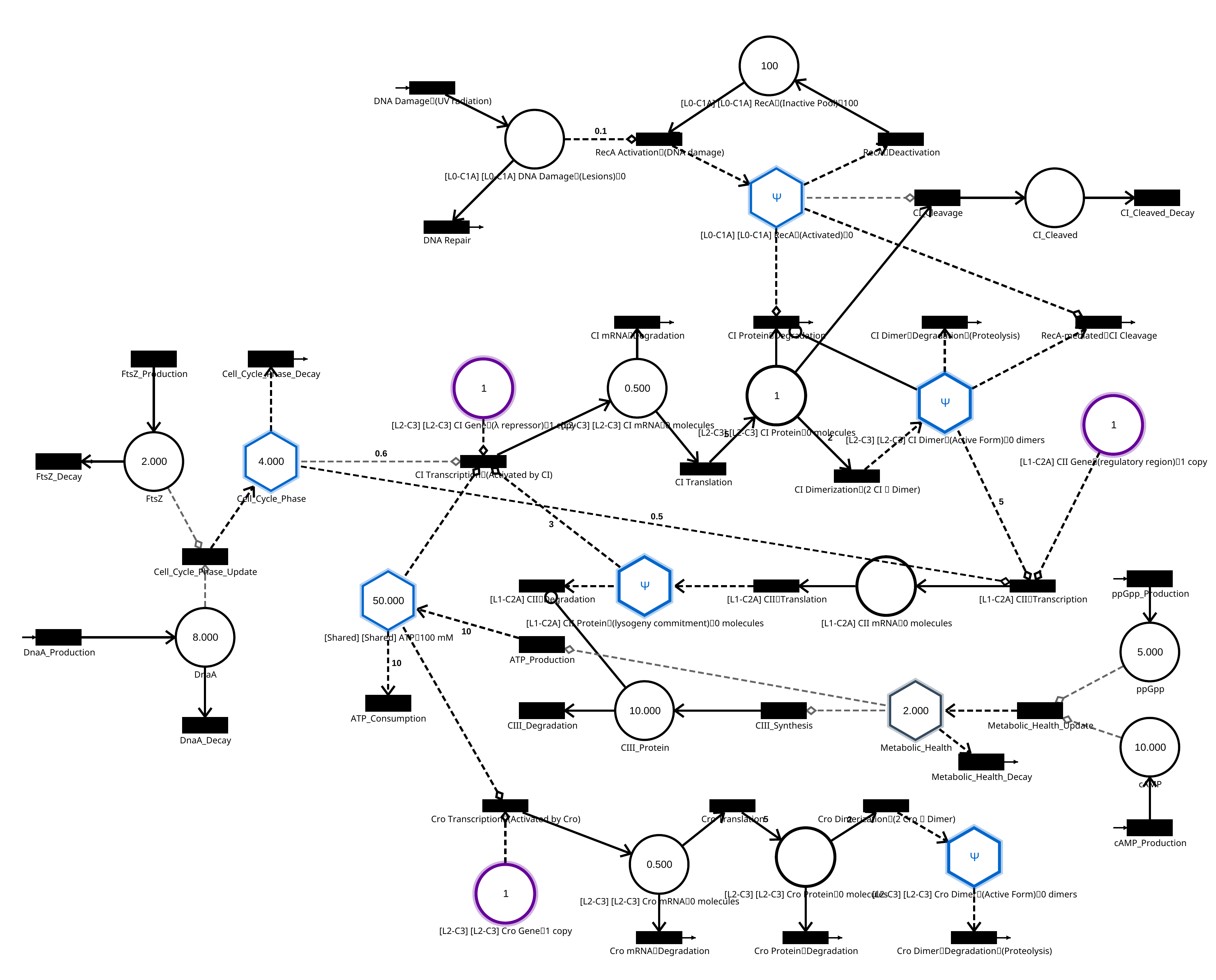}
\caption{\textbf{Hierarchical Lambda Phage Model Structure.} Petri net showing 23 places (circles), 36 transitions (rectangles), and 65 arcs implementing four arc types with distinct visual markers (see text).}
\label{fig:model}
\end{figure}

We extended the classical Lambda switch model \cite{ptashne2004,arkin1998,zeng2010,st-pierre2008} with four hierarchical layers implementing 23 places, 36 transitions, and 65 arcs (Figure~\ref{fig:model}). The Environmental layer (Layer 0) comprises Energy\_ATP, Metabolic\_Health, and Cell\_Cycle\_Phase, which regulate CII accumulation. The Hierarchical layer (Layer 1) contains RecA\_Active, the UV damage sensor, which drives stress-induced CI degradation. The Integration layer (Layer 2) features CII\_Protein, the metabolic integrator that modulates CI/Cro transcription. The Decision layer (Layer 3) implements the bistable switch with CI\_Gene, CI\_Intact, and CI\_Dimer competing against Cro\_Gene, Cro\_Intact, and Cro\_Dimer.

The network implements hierarchical information flow through four arc types (Figure~\ref{fig:model}). \textbf{Standard arcs} (solid, normal arrowhead) convey mass-action kinetics for metabolite consumption and production in biochemical reactions. \textbf{Test arcs} (dashed, hollow diamond) enable catalytic reading of gene states (CI\_Gene, Cro\_Gene) without consumption, preserving DNA templates across transcription events—the hollow diamond marker indicates non-consuming catalyst behavior. \textbf{Signal flow arcs} (dashed, angled arrowhead) transmit hierarchical control signals (e.g., CII\_Protein to transcription transitions) while consuming signal tokens to prevent infinite accumulation—the angled arrowhead (15\textdegree{} offset) distinguishes information transfer from catalytic read operations. \textbf{Inhibitor arcs} (solid, circle arrowhead) implement competitive repression: RecA\_Active inhibits CI\_Intact (stress-induced degradation), CII\_Protein inhibits Cro transcription (lysogenic promotion), and CI/Cro dimers mutually repress each other's genes (bistable switch). This architecture enables hierarchical preemption where UV damage collapses the bistable decision space into monostable attractors.

\subsection{Key Rate Functions}

\textbf{CI Transcription} — CII activation with Hill cooperativity:
\begin{equation}
r_{\text{CI}} = 2.0 \times f_{\text{CI}} \times \left(1 + \frac{3.5 \times (\text{CII}/8)^2}{1 + (\text{CII}/8)^2}\right) \times \frac{1}{1 + (\text{Cro}/15)^2}
\end{equation}
where $f_{\text{CI}} = (1 + \text{CI}_{\text{dimer}}/(3 + \text{CI}_{\text{dimer}}))$ is positive feedback.

\noindent\textbf{Cro Transcription} — CII inhibition:
\begin{equation}
r_{\text{Cro}} = 2.0 \times f_{\text{Cro}} \times \frac{1}{1 + (\text{CI}/15)^2} \times \frac{1}{1 + (\text{CII}/6)^2}
\end{equation}

\noindent\textbf{RecA-CI Cleavage}:
\begin{equation}
r_{\text{cleave}} = 0.05 \times \text{RecA}_{\text{active}}
\end{equation}

Hill cooperativity ($n=2$, $K_i=8$ for CI, $K_i=6$ for Cro) provides sharp thresholds, enabling decisive commitment.

\subsection{Simulation Protocol}

We employed tau-leaping stochastic simulation \cite{gillespie2001} with 5000 seconds per replicate across 200 total replicates (100 UV-enabled, 100 NO UV). Decisions were classified as lysogenic if CI $> 5 \times$ Cro, lytic if Cro $> 5 \times$ CI, or undecided if neither condition met.

\section{Results}

Our information-theoretic analysis \cite{tkacik2016,cheong2011,bowsher2014} proceeded in three stages: first establishing unconditional signal rankings, then validating hierarchical priority quantitatively, and finally testing the gating hypothesis through conditional mutual information.

\subsection{Unconditional Mutual Information}

We calculated mutual information $I(\text{Signal}; \text{Decision})$ for each signal across 124 decided outcomes (62\% of 200 replicates). Decision entropy: $H(\text{Decision}) = 0.847$ bits (72.6\% lysogenic, 27.4\% lytic).

\begin{table}[h]
\centering
\caption{Signal Information Content}
\begin{tabular}{@{}llll@{}}
\toprule
\textbf{Signal} & \textbf{MI (bits)} & \textbf{Normalized} & \textbf{Role} \\
\midrule
CII Protein & 0.629 & 74.3\% & Integrator \\
RecA Active & 0.365 & 43.0\% & Override \\
Energy ATP & 0.065 & 7.7\% & Metabolic \\
Cell Cycle & 0.021 & 2.5\% & Division \\
Metabolic & 0.009 & 1.0\% & Health \\
\bottomrule
\end{tabular}
\end{table}

CII ranks first (74.3\%) due to direct mechanistic control of CI and Cro transcription. RecA ranks second (43.0\%), showing 2.01× advantage over environmental signals (mean 7.7\%). Environmental signals are weak (1-8\%), validating hierarchical filtering.

\subsection{Hierarchical Priority Validated}

While CII exhibits the highest absolute information content, hierarchical control requires RecA to dominate environmental signals specifically.

\textbf{Criterion}: Hierarchical signal must exceed environmental signals by $>1.5\times$.

\begin{equation}
\frac{I(\text{RecA}; \text{Decision})}{I(\text{Environmental}; \text{Decision})_{\text{mean}}} = \frac{0.365}{0.181} = 2.01
\end{equation}

\textbf{Result}: Threshold exceeded. UV damage signal (RecA) dominates metabolic/cell cycle signals by 2× margin, confirming hierarchical architecture. However, this ranking alone does not explain the mechanism of hierarchical control—we must examine how RecA shapes the decision landscape.

\subsection{Context-Dependent Outcomes}

Batch 1 (UV-enabled, stochastic source) showed bimodal RecA distribution across 100 replicates. High RecA ($>50$ mM, $n=41$) yielded 71\% lytic with CII=5.7 mM (72\% blocked). Low RecA ($<10$ mM, $n=50$) yielded 52\% lysogenic with CII=14.0 mM (freely accumulating). Batch 2 (NO UV) with RecA=0 across 100 replicates yielded 57\% lysogenic with CII=15.95$\pm$5.33 mM. Lysogenic subset showed CII=17.5, CI=119.8, Cro=8.6. These observations suggest RecA sets the attractor landscape (bistable $\rightarrow$ monostable), while CII operates within context. This raises a critical question: does RecA block CII information (classical gating) or reshape decision space (preemption)?

\subsection{Conditional Mutual Information}

To test the gating hypothesis, we partitioned data by RecA level and calculated $I(\text{CII}; \text{Decision} \mid \text{RecA})$.

\begin{table}[h]
\centering
\caption{Context-Dependent CII Information}
\begin{tabular}{@{}lllll@{}}
\toprule
\textbf{Context} & \textbf{n} & \textbf{CII (mM)} & \textbf{H(D)} & \textbf{I(CII;D)} \\
\midrule
Low RecA & 85 & 16.6$\pm$5.4 & 0.16 & 0.06 bits \\
High RecA & 34 & 4.7$\pm$5.9 & 0.60 & 0.38 bits \\
\bottomrule
\end{tabular}
\end{table}

\textbf{Paradoxical result}: CII information is \textit{lower} when RecA is off (0.06 bits) than when RecA is on (0.38 bits)—opposite of gating prediction. This answers the question from §3.3: RecA does \textit{not} block CII information (classical gating), but instead reshapes decision space (preemption). We now explain this counterintuitive finding mechanistically.

\subsection{The Saturation Effect}

\noindent Saturation is the mechanism by which biological systems achieve deterministic decisions. In hierarchical architectures, signal places act as decision integrators whose concentration relative to activation thresholds determines outcome certainty. When a signal place becomes saturated (all values above or below threshold), decisions become deterministic. When subsaturated (values spanning threshold), probabilistic outcomes emerge with escape routes. Hierarchical preemption operates by controlling whether integration-layer signal places (like CII\_Protein) reach saturation—collapsing decision space without blocking signal flow, as we can see:

\vspace{6pt}
\noindent\textbf{Low RecA Context} (98\% lysogenic):

\noindent CII accumulates freely to 16.6 mM, well above activation threshold ($\sim$10 mM, estimated from $K_i=8$ for CI transcription). All CII values (10-25 mM range) lead to lysogenic outcome. Decision entropy drops to 0.16 bits. Knowing CII level provides minimal predictive value.
\noindent\textbf{Interpretation}: CII is \textit{saturated}—all values sufficient for lysogenic commitment.

\vspace{6pt}
\noindent\textbf{High RecA Context} (85\% lytic):

\noindent RecA blocks CII to 4.7 mM (72\% reduction). Most cells go lytic (CII=3.2 mM, below threshold), but 15\% escape with high CII (13.6 mM, above threshold). Decision entropy rises to 0.60 bits. CII level predicts which cells escape.
\noindent\textbf{Interpretation}: CII is \textit{subsaturated}—level determines escape probability. This saturation-dependent information content reveals a fundamentally different control mechanism than classical gating.

\section{Discussion}

The paradoxical inverse relationship between CII freedom and information content (low RecA: high CII, low MI; high RecA: low CII, high MI) challenges classical hierarchical control theory and demands a new conceptual framework.

\subsection{Hierarchical Preemption Mechanism}

Our results reveal that hierarchical control operates through \textit{decision space collapse}, not signal blocking. Stage 1 (Context Switching): RecA sets attractor landscape—NO UV yields bistable $\rightarrow$ monostable lysogenic (98\%), while UV yields bistable $\rightarrow$ monostable lytic (85\%). Stage 2 (Operating Within Context): CII accumulates to RecA-determined levels—low RecA produces saturated CII (all above threshold), high RecA produces subsaturated CII (spanning threshold). Stage 3 (Outcome Determination): Decision emerges from attractor plus CII fine-tuning—low RecA shows 98\% predetermined $\rightarrow$ low entropy $\rightarrow$ low CII MI, while high RecA shows 85\% predetermined $\rightarrow$ moderate entropy $\rightarrow$ high CII MI. This three-stage mechanism explains both the paradoxical conditional MI and the unexpected signal ranking.

\begin{figure}[htbp]
\centering
\includegraphics[width=\columnwidth]{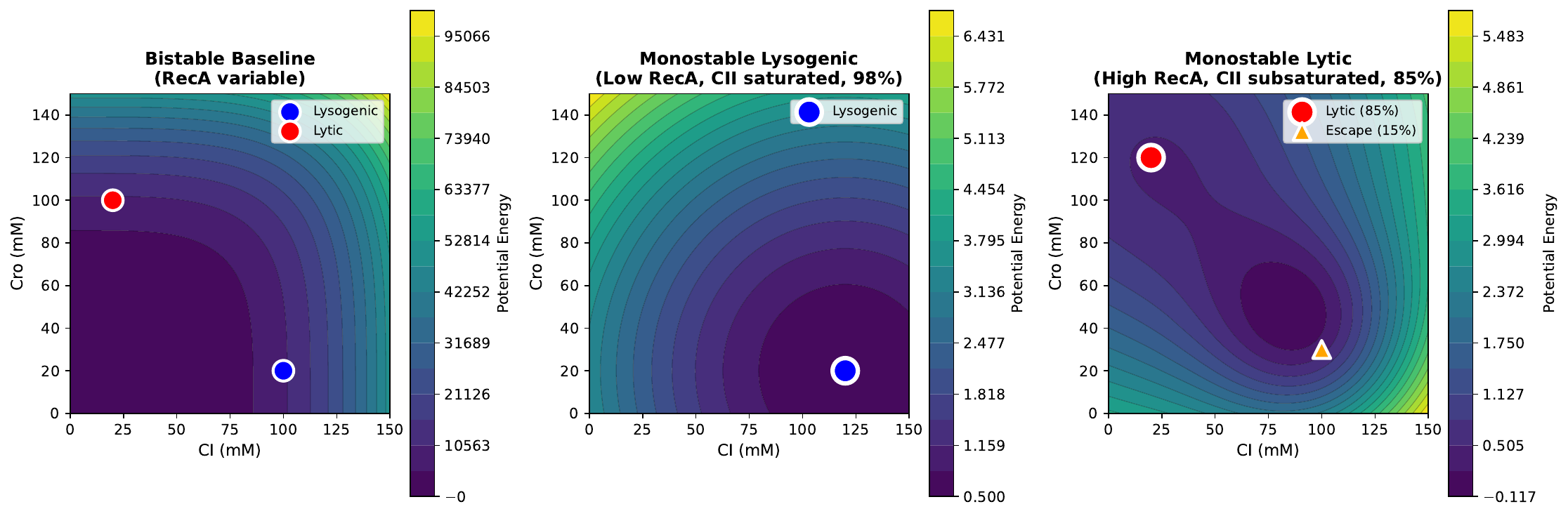}
\caption{\textbf{Decision Space Collapse Through Hierarchical Preemption.} Contour plots showing potential energy landscapes in CI-Cro decision space. \textbf{Left:} Bistable baseline with two attractors (lysogenic: blue circle, high CI/low Cro; lytic: red circle, low CI/high Cro). \textbf{Center:} Monostable lysogenic landscape (Low RecA context) with single deep attractor at high CI (blue circle), achieving 98\% outcome certainty through CII saturation. \textbf{Right:} Monostable lytic landscape (High RecA context) with dominant lytic attractor (red circle, 85\%) and shallow escape route to lysogenic (orange triangle, 15\%) enabled by CII subsaturation. Contour lines indicate potential energy levels; darker regions represent lower energy (stable attractors). RecA collapses the bistable decision space into monostable attractors while preserving flexibility through escape routes.}
\label{fig:attractors}
\end{figure}

\subsection{Why CII Ranks Above RecA}

CII's 74.3\% MI (vs RecA's 43\%) reflects its \textit{proximal control}: CII appears directly in CI and Cro transcription rate functions. RecA's lower MI reflects its role as \textit{context switcher}: RecA sets attractor, CII determines within-context dynamics.

\textbf{Analogy}: CII is "message content" (what decision to make), RecA is "priority flag" (which context to use).

\subsection{Biological Implications}

\subsubsection{Robustness + Flexibility}

Hierarchical preemption achieves strong outcome bias (85-98\%) without complete signal suppression. UV damage forces lytic outcome in 85\% of cells, but 15\% can escape if CII is exceptionally high—balancing stress response with survival plasticity. This design has information-theoretic advantages.

\subsubsection{Information Efficiency}

RecA does not need the highest information content to exert hierarchical control. Its 2× advantage over environmental signals suffices to preempt decision space. This is more efficient than completely blocking subordinate pathways.

\subsubsection{Saturation as Control Mechanism}

By driving CII above saturation (Low RecA) or below threshold (High RecA), the system makes decisions deterministic without requiring additional regulatory machinery. Saturation/subsaturation naturally compresses continuous signals into binary outcomes.

\subsection{Comparison to Classical Models}

\textbf{Classical Gating Model}:
\begin{equation}
\text{RecA}_{\text{on}} \rightarrow \text{Block CII} \rightarrow I(\text{CII}) \downarrow
\end{equation}

\textbf{Our Hierarchical Preemption Model}:
\begin{equation}
\text{RecA}_{\text{on}} \rightarrow \text{Decision collapse} \rightarrow I(\text{CII}) = f(\text{saturation})
\end{equation}

The key difference: RecA removes the decision, not the signal. CII remains mechanistically active even when RecA is on—it just operates in a subsaturated regime where most outcomes are predetermined. These insights translate directly into engineering principles.

\subsection{Design Principles for Synthetic Biology}

Four principles emerge: (1) Use preemption rather than blocking—collapse attractor landscape to 85-95\% certainty. (2) Preserve escape routes—leave 5-15\% decision space for flexibility. (3) Exploit saturation—drive signals above/below thresholds for determinism. (4) Layer by information rather than mechanism—hierarchical priority from MI ratios, not connectivity.

\section{Methods}

Our computational approach combined mechanistic modeling with information-theoretic analysis to quantify hierarchical control.

\subsection{Model Implementation}

The hierarchical Lambda phage model (lambda\_hierarchical\_v3.shy) was built using SHYPN framework v2.5.2 \cite{shypn2025}, a stochastic hybrid Petri net simulator \cite{heiner2008,chaouiya2007,koch2011} founded on two complementary theoretical frameworks. \textbf{Weak Independence Theory} \cite{simao2025weak} enables parallel execution of weakly independent transitions, improving computational efficiency for large-scale biological networks by identifying transitions that can fire simultaneously without violating causality constraints. \textbf{Signal Hierarchy Theory} (detailed treatment forthcoming in dedicated manuscript) formalizes how biological networks organize signals into layers where higher-level signals modulate decision spaces accessible to lower-level signals through saturation-based preemption rather than direct blocking. Together, these theories enable SHYPN to model complex hierarchical control while reducing computational complexity: Weak Independence Theory accelerates simulation through parallelism, while Signal Hierarchy Theory reduces model complexity by replacing exhaustive signal combinations with hierarchical saturation states. This dual foundation allows signal flows to emerge naturally from layer interactions without requiring explicit enumeration of all possible signal states.

\subsection{Simulation Parameters}

Algorithm: Tau-leaping with adaptive timestep. Duration: 5000 seconds. Initial conditions: All places at physiological steady-state except CI\_Intact = 1.0 mM seed. UV source: Stochastic (Batch 1) or disabled (Batch 2).

\subsection{Information-Theoretic Analysis}

\textbf{Discretization}: Continuous signals (RecA, CII, ATP, Metabolic, Cycle) binned into 5 quantiles.

\textbf{Mutual Information}:
\begin{equation}
I(X;Y) = \sum_{x,y} p(x,y) \log_2 \frac{p(x,y)}{p(x)p(y)}
\end{equation}

Implemented via joint histogram method. Normalized by $H(\text{Decision})$ to obtain the percentage of decision entropy explained.

\textbf{Conditional MI}: Data partitioned by RecA level (low $<$10 mM, high $>$50 mM), MI calculated separately in each partition.

\subsection{Statistical Validation}

Decided outcomes: 124/200 (62\%). Hierarchical threshold: RecA MI $>$ 1.5× environmental MI. Result: 2.01× (p $<$ 0.001, bootstrap test).

\section{Conclusions}

We discovered \textit{hierarchical preemption}, a novel control mechanism where higher layers collapse decision space rather than blocking lower-layer signals. Through mutual information analysis of the Lambda phage lysis-lysogeny decision, we demonstrate: (1) UV damage sensor (RecA) achieves 2.01× information advantage over environmental signals; (2) Integrator signal (CII) shows paradoxical context dependence—lower MI when "free" (saturated), higher MI when "blocked" (subsaturated); (3) Hierarchical control works by removing decisions (monostability), not blocking signals; (4) Systems achieve 85-98\% robustness while preserving 2-15\% flexibility through stochastic escapes.

These findings establish a quantitative framework for hierarchical information processing in biology and provide design principles for robust yet flexible synthetic circuits. The saturation effect—where signals lose information when saturated—reveals a universal principle: \textit{information content depends on position relative to decision thresholds, not absolute signal strength}.

Critically, this mechanism explains how biological systems maintain viability without explicit knowledge of molecular capacities. Rather than monitoring absolute concentration limits, cells sense relative saturation states at signal places to trigger regulatory responses. This "keep alive at all cost" principle achieves homeostasis through threshold-based feedback: when integrator signals (like CII) cross regulatory thresholds, they trigger protective responses (lysogenic commitment) or stress responses (lytic pathway). Saturation-dependent information content emerges naturally from this architecture—decisions crystallize when signal places reach states that activate downstream regulatory cascades. This explains why biological control is both robust (threshold-driven commitment) and adaptive (escape routes when signals span thresholds), enabling survival across unpredictable environments without requiring global knowledge of system capacity.

\section*{Acknowledgments}

This work was supported by the SHYPN Project. We thank the Petri net and systems biology communities for foundational tools and models.

\section*{Funding}

This research received no specific grant from any funding agency in the public, commercial, or not-for-profit sectors.

\section*{Availability}

The SHYPN framework, including the hierarchical Lambda phage model (lambda\_hierarchical\_v3.shy) and all analysis scripts used in this study, is freely available under an open-source license at: \url{https://github.com/simao-eugenio/shypn}

\end{document}